\documentclass[runningheads]{llncs}

\usepackage{eccv}

\usepackage{eccvabbrv}

\usepackage{graphicx}
\usepackage{booktabs}

\usepackage{algorithm,algorithmic}
\usepackage{multirow}
\usepackage{tabularx}
\usepackage{makecell}
\usepackage{colortbl}

\usepackage{pifont}
\usepackage{lipsum}
\usepackage[accsupp]{axessibility}  % Improves PDF readability for those with disabilities.

\usepackage{hyperref}

\usepackage{orcidlink}

\begin{document}

\title{CipherDM: Secure Three-Party Inference for Diffusion Model Sampling} 

\titlerunning{CipherDM: Secure Three-Party Inference for Diffusion Model Sampling}

\author{Xin Zhao\inst{1,2}\orcidlink{0009-0005-5730-7534} \and
Xiaojun Chen\inst{1,2}\orcidlink{0000-0003-0362-847X} \and
Xudong Chen\inst{1,2}\orcidlink{0009--0007-0369-9899} \and
He Li\inst{1,2}\orcidlink{0000--0003-0611-3236}
\and
Tingyu Fan\inst{1,2}\orcidlink{0000--0002-1530-1928}
\and
Zhendong Zhao\inst{1,2}\orcidlink{0000--0003-0003-019X}
}

\authorrunning{X.~Zhao et al.}


\institute{School of Cyberspace Security, University of Chinese Academy of Sciences, China \and
Key Laboratory of Cyberspace Security Defense, Institute of Information Engineering, Chinese Academy of Sciences,
China
\email{\{zhaoxin,chenxiaojun,chenxudong,lihe2023,fantingyu,zhaozhendong\}@iie.ac.cn}\\
}

\maketitle
\begin{abstract}
   Diffusion Models (DMs) achieve state-of-the-art synthesis results in image generation and have been applied to various fields. However, DMs sometimes seriously violate user privacy during usage, making the protection of privacy an urgent issue. Using traditional privacy computing schemes like Secure Multi-Party Computation (MPC)  directly in DMs faces significant computation and communication challenges. To address these issues, we propose CipherDM, the first novel, versatile and universal framework applying MPC technology to DMs for secure sampling,  which can be widely implemented on multiple DM based tasks.  We  thoroughly analyze sampling latency breakdown, find time-consuming parts and design corresponding secure MPC protocols for computing nonlinear activations including SoftMax, SiLU and Mish. CipherDM is evaluated on popular architectures (DDPM, DDIM) using MNIST dataset and on SD deployed by diffusers. Compared to direct implementation on SPU, our approach improves running time by approximately $1.084\times \sim 2.328 \times$, and reduces communication costs by approximately $1.212\times \sim 1.791 \times$. \\ Code is available at: \href{https://github.com/Zhaoxinxinzi/CipherDM}{https://github.com/Zhaoxinxinzi/CipherDM}.
\end{abstract}

\section{Introduction}
\label{sec:intro}
    Generative models represented by Pre-trained Transformer Models\cite{Attention} and Diffusion Models (DMs)\cite{DDPM,2015deep,Scorebased} have gained significant attention due to their impressive performance in text and image generation tasks. These models have become widely utilized in Deep Learning as a Service (DLaaS) paradigm\cite{DLaaS}. However, the utilization of such services raises concerns about privacy.  In the case of ChatGPT \cite{ChatGPT} and Stable Diffusion (SD) \cite{Stable}, users are required to disclose their private prompts or images to service providers, such as websites or apps that possess substantial computing resources. Alternatively, service providers may release their proprietary trained model parameters to users, who can then perform inference or sampling locally.
    
    The remarkable performance of DMs in image synthesis has been demonstrated in several recent works\cite{Stable, Glide, Dalle2, Dalle,easyphoto}. However, these works also raise concerns about privacy issues\cite{Stable, Glide, Dalle2}. For instance, the fundamental work of SD\cite{Stable} claims that training data used in DMs may contain sensitive or personal information, and there might be a lack of explicit consent during data collection process.  On the other hand, Glide\cite{Glide}
    emphasizes that their model has the capability to generate fake but highly realistic images, which could potentially be used to create convincing disinformation or Deepfakes.  To address ethical concerns, Glide filters out training images that contain people, violence, and hate symbols, thereby reducing model's potential for misuse in problematic scenarios. Another example is the work by Ramesh\cite{Dalle2}, which trains their model on a specific dataset that is carefully filtered to ensure aesthetic quality and safety. These works primarily focus on issues of data privacy and content realism. Despite implementing measures to mitigate safety concerns, they cannot guarantee the protection of privacy information sufficiently. 
    
    Traditional sampling phase \cite{DDPM,DDIM, kdiffusion, zhao2023unipc} faces risk of leaking users' private prompts or images, as well as providers' model parameters. For users, the prompts and images used in sampling process may contain sensitive information, including personal portrait, identities, habits and professions. The unauthorized collection of such information by illegal organizations poses a significant threat to user privacy. Similarly, for model providers, the training process of DMs requires substantial investments in terms of time and resources. The leakage of model details, including parameters and architecture, can significantly impact their benefits and competitive advantage. Given these concerns, it is crucial to prioritize the protection of privacy during sampling process. 

    Classic privacy computing categorizes to Differential Privacy (DP)\cite{DP1,DP2,DP3,DP4}, Homomorphic Encryption (HE) \cite{HE1,HE2,HE3,HE4} and Secure Multiparty Computation (MPC) \cite{ABY,ABY2,ABY3,Meteor,secureML,miniONN}. Although DP is generally considered as an efficient approach, it cannot effectively guarantee precision and accuracy of the calculation results.  Additionally, it currently lacks rigorous security proofs. HE maintains confidentiality of data at the expense of high computational complexity. MPC offers a solution that allows mutually distrusting parties to collaboratively evaluate a function using their private inputs, while ensures that only output of the function is revealed and no party can gain additional information. This computation guarantees both privacy and correctness, preventing corrupt parties from learning anything but the output and ensuring that honest parties do not accept an incorrect output. Several works\cite{MPCformer,CipherGPT,Puma,mpcbert,mpcvit} have proposed MPC-based solutions for performing inference on transformer models. However, to the best of our knowledge, no MPC-based approaches for secure sampling on DMs have been developed thus far. To our conjectures, the primary reasons for absence of related work may contain unclear privacy assessment in image generation results, difficulty in integrating DMs with MPC frameworks and low sampling efficiency in private computing.
    
    Based on aforementioned observations, we present CipherDM, the first MPC-based inference framework that provides rigorous MPC guarantees for sampling DMs using the ABY3 protocol\cite{ABY3}.  We employ a 3PC (Three-Party Computation) \cite{Abadi_2016,ABY3} replicated secret sharing scheme in an outsourced scene. In this scheme, enquirer shares their private data, while model owner shares model parameters, with three separate cloud servers. The servers then collectively perform sampling phase and return reshared result to the enquirer. Throughout this entire process, neither private input data nor model parameters will be exposed to other parties. 
\begin{figure}[ht]
  \centering
    \includegraphics[height=2.7cm]{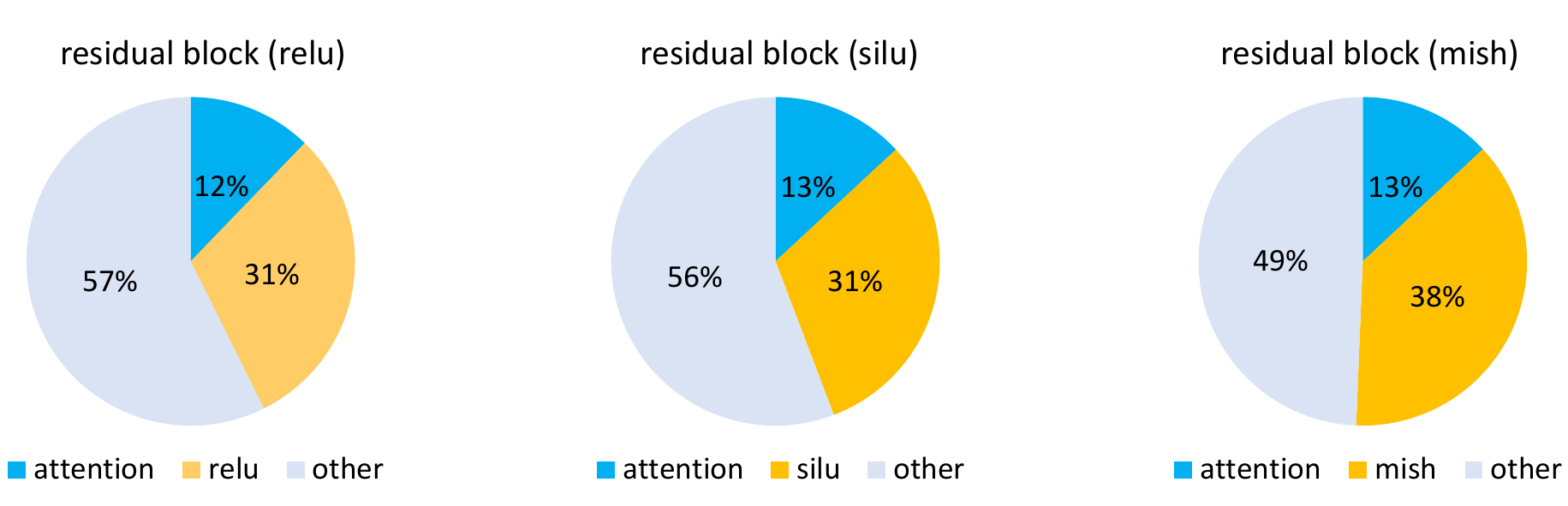}
  \caption{ Module running time percentage of residual block in plaintext. 
  }
  \label{fig:residual}
\end{figure}

    We measure sampling running time in plaintext and discover that the dominant factor influencing overall cost is noise prediction step, which relies on estimation provided by the U-Net \cite{unet} module. The result shown as \cref{fig:residual} indicates that  residual block is the most time-consuming component, accounting for approximately 82.87\% of  total noise prediction time. Furthermore, the attention block and activation layers together take up nearly half of the time consumed by residual block.  

    \begin{figure}[ht]
      \centering
    \includegraphics[height=3.3cm]{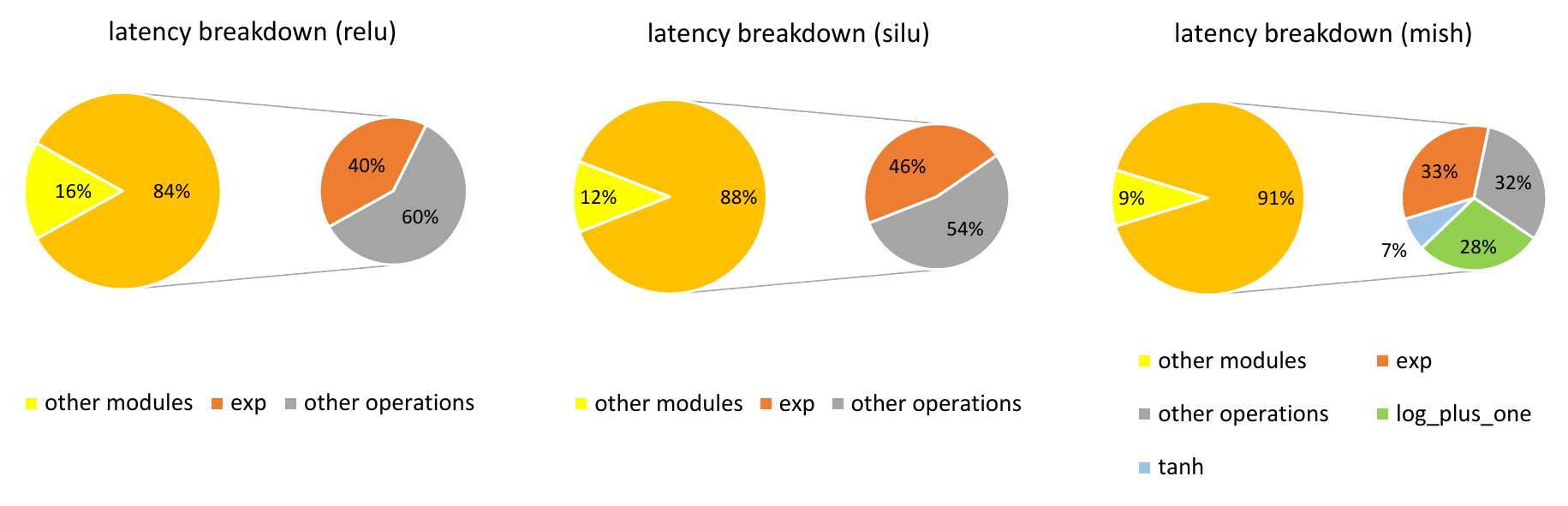}
      \caption{ Latency breakdown of total workflow in ciphertext.
      }
      \label{fig:lantency}
    \end{figure}
    
    Moreover, the ciphertext profiling presented in \cref{fig:lantency} reveals that the majority time in sampling process is occupied by MPC phase.  The main graphs (large circle of each) show total running time under SPU\cite{spu}. The orange blocks represent MPC execution time and the yellow blocks represent time for other modules including setup and compiling. The subgraphs (small circle of each) demonstrate the percentage of time that nonlinear operations perform on MPC. Exp is short for exponentiation, tanh is short for hyperbolic tangent and log\_plus\_one is the logarithmic item of softplus function. The grey blocks refer to other operations including add, multiplication, division, \etc. It is obvious that nonlinear operations serve as the primary bottleneck within MPC phase. According to these analysis, we optimize the computation through approximating complex nonlinear operations above  with linear polynomials and design corresponding secure protocols. 
    
    Our contributions can be summarized as follows:
    
    \textbf{Secure DM Sampling:} We first leverage MPC technology to ensure confidentiality and privacy of the sampling phase. Specifically, we implement a 3PC replicated secret sharing scheme in an outsourced setting. We successfully generate one image within 10 minutes under secret mode at best. Given that both sampling DMs and MPC are known to be time-consuming processes, the combination of these two approaches is often considered challenging to achieve. Consequently, the impressive results obtained from their successful integration are quite surprising.

    \textbf{Unified Secure DM:} We confirm that MPC can be effectively implemented on commonly used DMs. To assess the performance, we conduct experiments on several well-known DM algorithms, including DDPM\cite{DDPM}, DDIM\cite{DDIM} and SD\cite{Stable}. This validation process helps us determine the practicality and potential benefits of applying MPC to securify sampling phase of DMs.

    \textbf{Optimized Secure Nonlienar Operators:} We develop efficient activation protocols specifically designed for DMs. These protocols are tailored to handle SoftMax activation, which is essential in DMs, as well as optional activations such as ReLU, SiLU, and Mish\cite{mish}. Compared to directly implemented on SPU\cite{spu}, we achieve approximately $1.084\times \sim 2.328\times$ improvement in terms of computation time, and $1.212\times \sim 1.791\times$ reduction in communication costs.

\section{Related Work}

 \textbf{Diffusion Models}\ DMs have gained prominence and  surpassed the prolonged dominance of Generative Adversarial Networks (GANs) \cite{GAN}, emerging as the new state-of-the-art in generative modeling. They have demonstrated exceptional performance in a wide range of applications, including image synthesis\cite{cdm,Stable,Dalle2,easyphoto,Glide}, video generation \cite{vdm,dit,fdm} and other fields\cite{3dpoint,anomaly,junction,diffusionlm}, setting new records in terms of quality and realism. In this work, we evaluate on several popular architectures including DDPM\cite{DDPM}, DDIM\cite{DDIM} and SD\cite{Stable}.

\textbf{Privacy Protection}\  Tim Dockhorn $et$ $al.$ \cite{DPDM} utilizes Differential Privacy (DP) to protect sensitive data in the training phase of DMs. Xiao He $et$ $al.$ \cite{Diffprivacy} unifies anonymization and visual indentity information hiding to achieve face privacy protection based on SD\cite{Stable}. Diffence\cite{diffence} employs DMs as a defense mechanism against member inference attacks. Existing works primarily address data privacy \cite{DPDM,Stable}, content legacy\cite{Dalle2} and  content reality\cite{Glide}, lacking concern on input and model privacy for DM sampling. Privacy-Preserving Machine Learning (PPML)\cite{privacypreserving,DP3} has done substantial inference work in machine learning related tasks and shown promise in various applications including healthcare, vehicular networks, intelligent manufacturing, \etc.  They have made great process in protecting privacy, improving efficiency and reducing communication\cite{ABY2,miniONN,HE3,delphi,secureML}. Therefore, PPML related technologies are possible to be transferred to Privacy-Preserving Diffusion Model.

 \textbf{Secure Multiparty Computation}\  MPC technologies used for privacy inference mainly include Secret-sharing (SS) \cite{ABY, ABY2,ABY3, miniONN, secureML} and Homomorphic Encryption (HE) \cite{HE1,HE2,HE3,HE4}. The architecture of MPC can be classified into two-party setting\cite{miniONN,secureML,delphi,Cheetah,CrypTFlow2,ABY2}, three-party setting\cite{ABY3,securenn,Meteor,CrypTFlow}, four-party setting\cite{FLASH,Fantastic} and \etc\cite{MOTION}.
Among these works, three-party setting has attached much attention because it has the highest concrete efficiency in resisting semi-honest adversaries in honest majority.  Existing research works only consider secure inference of PPML\cite{miniONN,secureML,delphi,Cheetah} and Transformer models\cite{Puma,MPCformer,mpcvit,mpcbert}. However, there is no related work concerning DM sampling so far. That is to say, our CipherDM is the groundbreaking work and has a significant impact.

\section{Background}

\subsection{Diffusion Models}
Diffusion Models (DMs) utilize a two-step Markov chain process to generate image samples from initial noise images. This process can be divided into two distinct chains: the \textit{forward process} and the \textit{reverse process}. The \textit{forward process} in DMs can be likened to a Brownian motion, where a real image $x_0$ is transformed gradually into a latent Gaussian noise space $x_T$.  This procedure tends to model data distribution by approximating the intermediate images between $x_0$ and $x_T$. Besides, the \textit{reverse process}, also known as the backward Markov chain, aims to generate the initial image $x_0$ by leveraging a learned Gaussian transition. Its goal is to infer the initial image distribution conditioned on the final image. By combining both forward and reverse processes, DMs can effectively capture the complex data distribution and generate high-quality samples.

DDPM\cite{DDPM} is a representative diffusion model that motivates many follow-up works. To explain how CipherDM inference on DMs, we take DDPM for example and provide a brief review of its underlying mechanism.  DDPM maps Gaussian distribution $\mathcal{N}(x_T;0,\textbf{I})$ to the distribution of real images $q(x_0)$, and aims to recover initial $x_0$ from the mapped noise image. 

The \textit{forward process}  gradually adds  Gaussian noise to the data sample according to the variance schedule $\beta_1, ..., \beta_T$ and finally reaches a standard Gaussian distribution $x_T \sim \mathcal{N}(0, \textbf{I})$. Because of the well-designed variance schedule, we can express $x_t$ at any arbitrary timestep $t$ in closed form \cref{eq:qnoise} with the notation $\alpha_t:=1-\beta_t$ and $\Bar{\alpha}_t:=\prod^t_{s=1}\alpha_s$. 

\begin{equation}
    q(x_t|x_0)=\mathcal{N}(x_t;\sqrt{\Bar{\alpha_t}}x_0,(1-\Bar{\alpha}_t)\textbf{I})
    \label{eq:qnoise}
\end{equation}

The \textit{reverse process} starts at $p(x_T) =\mathcal{N}(x_T;0,\textbf{I})$. Strict inference and proof conclude that the distribution of $x_{t-1}$ is also Gaussian given conditions of $x_0$ and $x_t$. Therefore, the  $x_{t-1}$ at any arbitrary timestep $t$ can also be expressed as \cref{eq:qposterior} where $x_t(x_0,\epsilon)=\sqrt{\Bar{\alpha}_t}x_t+\sqrt{1-\Bar{\alpha}_t}\epsilon$ for $\epsilon\sim\mathcal{N}(0,\textbf{I})$. Here $\Tilde{\mu}_t$ and $\Tilde{\beta}_t$ represent mean and variance of reverse Gaussian distribution respectively.

\begin{equation}
    \begin{split}  
    q(x_{t-1}|x_t,x_0)&:=\mathcal{N}(x_{t-1};\Tilde{\mu}_t(x_t,x_0),\Tilde{\beta}_t\textbf{I}) \quad where\\
    \Tilde{\mu}_t(x_t,x_0):=\frac{1}{\sqrt{\alpha_t}}(x_t(x_0&,\epsilon)-\frac{\beta_t}{\sqrt{1-\Bar{\alpha}_t}}\epsilon) \quad and \quad \Tilde{\beta}_t:=\frac{1-\Bar{\alpha}_{t-1}}{1-\Bar{\alpha}_t}\beta_t
    \end{split}
    \label{eq:qposterior}
\end{equation}

The sampling process corresponds to the reverse process of DMs. This procedure involves iteratively refining an initial noise-corrupted image to generate a high-quality image. The process unfolds over a series of steps, with each step gradually reducing noise and refining the image.
Each step is typically guided by a neural network, commonly U-Net\cite{unet}, which predicts noise to be subtracted from the image.

\subsection{2-out-of-3 Replicated Secret Sharing}
The main notations related to secret sharing and protocols are summarized in  \cref{tab:Notation}.

\begin{table}[t]
  \caption{Notation table.}
  \label{tab:Notation}
  \centering
  \scalebox{0.8}{
  \begin{tabular}{c|c}
   \toprule 
    \midrule
    $\mathbb{Z}_{2^{\ell}}$ & discrete ring modulo $2^{\ell}$\\
    
    $P_i$ & server $i$ in 3PC\\
    
    $[\![\cdot]\!]$ & Arithemetic Sharing in $\mathbb{Z}_{2^{\ell}}$ \\
    
   $[\![\cdot]\!]^B$ & Boolean Sharing in $\mathbb{Z}_{2}$  \\ 

   $\textbf{x}$ & uppercase bold letter denotes vector\\

    $x$ & lowercase letter denotes scalar \\

    $x^i$ & the power of scalar $x$ with an order of i \\

    $\textbf{t}^i$ & the power of vector $t$ with an order of i \\

    $\alpha_i$ & random value pre-computed by $P_i$\\

    $\oplus$ & the operation of XOR\\
   
 \toprule
  \bottomrule
\end{tabular}
    }
\end{table}

In 3PC setting, a secret value $x\in\mathbb{Z}_{2^l}$ is shared by  three random values $x_0,x_1,x_2\in\mathbb{Z}_{2^l}$ which satisfies $x=x_0+x_1+x_2$. Particularly in the 2-out-of-3 replicated secret sharing (denoted as $[\![\cdot]\!]$-sharing), party $P_i$ gets $[\![x]\!]_i = (x_i, x_{i+1})$. Without special declaration, we compute in $x\in\mathbb{Z}_{2^l}$ and omit (mod $2^l$) for brevity. We use notation $[\![\cdot]\!]$ to represent \textit{Arithemetic Sharing}, which supports arithemetic operations ($e.g., +,-$ and $\cdot$) when $l>1$ ($e.g., l=64$). While in the case of $l=1$ where ($+,-$) and $\cdot$ are respectively replaced by bit-wise $\oplus$ and $\wedge$, we refer to this type as \textit{Boolean Sharing}  ($[\![\cdot]\!]^B$).

\textbf{Addition}. Assuming  $(c_1,c_2,c_3)$ be public constants and $([\![x]\!],[\![y]\!])$ be two secret-shared values, we can compute  $[\![c_1x+c_2y+c_3]\!]$ through summarizing each part of  $(c_1x_0+c_2y_0+c_3, c_1x_1+c_2y_1, c_1x_2+c_2y_2)$, which can be calculated by  $P_i$ locally. Typically  we can get $[\![x+y]\!]$ by setting $c_1=1,c_2=1,c_3=0$ .

\textbf{Multiplication}. Given two shared values $[\![x]\!]$ and $[\![y]\!]$, the share $[\![x\cdot y]\!]$ can not be calculated by multipliating $[\![x]\!]$ and $[\![y]\!]$ directly. In the secure multiplication protocol $\prod_{\rm{Mul}}$, $P_i$ first computes $z_i = x_iy_i+x_{i+1}y_i+x_iy_{i+1}$ locally and sends $z_i'=\alpha_i+z_i$ to $P_{i-1}$. $\alpha_i$ is generated by $P_i$ in the setup phase as \cite{ABY3} and conform $\alpha_0+\alpha_1+\alpha_2=0$. Then $[\![x\cdot y]\!]$ can be formed by $\{(z_0',z_1'),(z_1',z_2'),(z_2',z_0')\}$ . 

\textbf{Underlying Protocols}. Apart from addition and multiplication, CipherDM also relies on several other underlying protocols. The underlying protocols are the same as implemented in Puma\cite{Puma} from the state-of-the-art work. They are employed in a blackbox manner and only enumerate the inputs and outputs as \cref{tab:Protocol}.

\begin{table}[t]
\setlength{\abovecaptionskip}{0cm}
 
  \caption{Underlying black-box secret-shared protocols.}
  \label{tab:Protocol}
  \centering
  \scalebox{0.8}{
  \begin{tabular}{l|l|l}
   \toprule   
    \midrule
    Protocol & Semantic & Expression \\
    \hline
   $\prod_{\rm{LT}}$ &  $z=1\{x<y\}$ & Less Than \\
   $\prod_{\rm{Max}}$  & $z=maximum(\textbf{x})$ & Maximum \\
   $\prod_{\rm{Recip}}$  & $z=1/x$& Reciprocal\\
   $\prod_{\rm{Square}}$  & $z=x^2$&Square\\
   $\prod_{\rm{Mul_{BA}}}$ & $z=b\cdot x$& Boolean-Arithmetic Multiplication \\
    
 \toprule
  \bottomrule
\end{tabular}
}
 
\end{table}

\textbf{Threat Model}. Following previous works \cite{ABY3}, CipherDM is secure against a semi-honest adversary that corrupts no more than one of the three computing parties. Semi-honest means such an adversary will observe protocol specifications, but may attempt to gain knowledge of other parties' private information during the execution of related protocols.

\textbf{Security Proof}. Given that our work builds upon several acknowledged fundamental works (\ie ABY3, SPU ) which have strict security proof and analysis regarding privacy preservation, and considering that we do not alter the underlying execution mechanism, we can confidently assert that our security/privacy is sufficiently guaranteed.

\section{Secure Design of CipherDM}
In this section, we first present an overview of CipherDM for securely sampling Diffusion Models (DMs) in \cref{sec:overview}.  Then we introduce  details of the Secure SoftMax protocol in \cref{sec:softmax}, the Secure Activations (SiLU and Mish\cite{mish}) protocol in \cref{sec:activa} and the Secure Time Embedding protocol in \cref{sec:temb}.

\subsection{Overview}
\label{sec:overview}
\begin{figure}[ht]
  \centering
    \includegraphics[height=5.0cm]{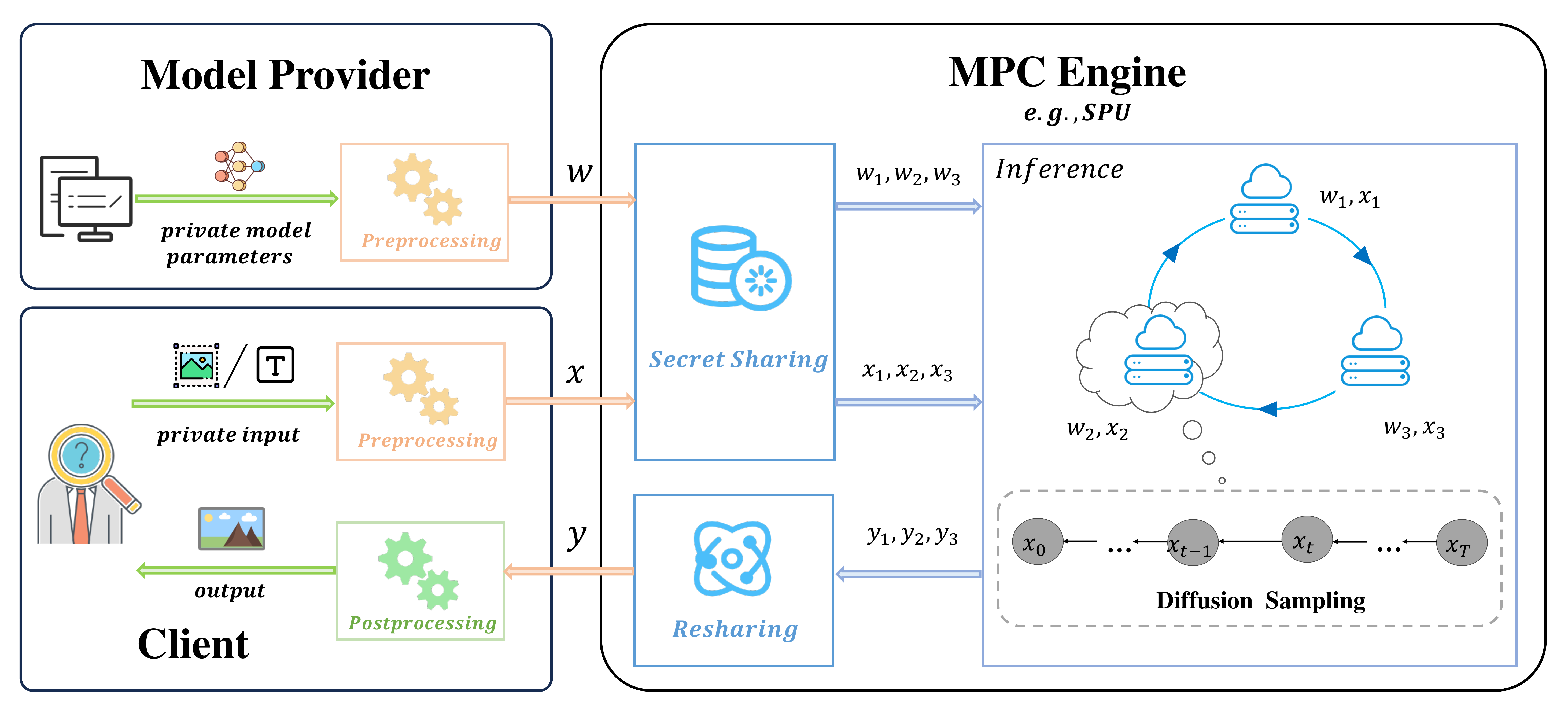}
  \caption{ An illustration of our proposed CipherDM framework. CipherDM takes model parameters and images/texts as two private inputs, preprocesses them locally, secretly shares them to a three-party MPC Engine, and receives the final sampling result from it.  MPC systems such as SPU involve  the joint computation.
  }
  \label{fig:overview}
  
\end{figure}

CipherDM adopts a secure outsourcing computing scenario where users send their private model and input to cloud servers to attain inference result. This situation makes sense in practice, because it is easy to find a credible third-party regulator that can provide computing power.  In CipherDM shown as \cref{fig:overview}, a model provider $\mathcal{M}$ holds private diffusion model parameters $w$, and the client $\mathcal{C}$ holds data $x$ (\eg, images or texts). $\mathcal{C}$ sends its private $x$ to cloud servers $\mathcal{S}$ after preprocessing locally, while $\mathcal{M}$ takes $w$ as input. Diffusion model architecture is typically considered public and accessible to all participants. Then $\mathcal{S}$ leverages an MPC engine to convert the inputs into secret sharing and distribute the sharing to three noncolluding parties, securely perform sampling process and return reshared output to $\mathcal{C}$. After postpocessing, $\mathcal{C}$ obtains final generated image. The preprocessing and postprocessing phases typically involve input/output and model-independent operations that are necessary for data preparation and result interpretation. These operations include tasks such as converting texts or images into embeddings, data normalization, and image storage. During the whole process, $\mathcal{C}$ gets nothing but the final result. Besides, neither $\mathcal{M}$ and $\mathcal{S}$ know $\mathcal{C}$'s input, nor $\mathcal{S}$ and $\mathcal{C}$ know model parameters.

\subsection{Secure SoftMax}
\label{sec:softmax}

The SoftMax function is employed in the attention block of U-Net\cite{unet}. It can be expressed as $\rm SoftMax(\textbf{x}[\textit i])= \frac{exp(\textbf{x}[\textit i]-\Bar{x}-\epsilon)}{\sum_i exp(\textbf{x}[\textit i]-\Bar{x}-\epsilon)}$, where $\Bar{x}$ is the maximum element of input vector $\textbf{x}$. $\epsilon$ is a tiny and positive value (\eg, $\epsilon=10^{-6}$) for ciphertext and equals 0 for plaintext. As the exponentiation function takes most time of computation shown as \cref{fig:lantency}, we replace it with Chebyshev polynomial\cite{chebyshev}, which can be computed as \cref{eq:exp}.
\begin{equation}
    \rm{negExp}(\textit x)=
    \begin{aligned}
    \begin{cases}
    0, &\ x<T_{exp} \\
    Chebyshev(x), &\ x\in[T_{exp},0] \\
    \end{cases}
    \end{aligned}
    \label{eq:exp}
\end{equation}

Supposing MPC system uses 18-bit fixed-point precision, we set $T_{exp}=-14$ given $exp(-14)<2^{-18}$, and then fit the approximating with a maximum order of 7.  The choice of this polynomial order is crucial in achieving an accurate approximation. Higher-order polynomials can capture more intricate details of the exponential function but may lead to increased computational complexity. On the other hand, lower-order polynomials may not capture the exponential behavior accurately.  Accordingly,  the computation of  $Chebyshev(x)$ can be expressed using Chebyshev polynomial as \cref{eq:cheb}. Here $x_t$ represents the mapping of $x$ from the interval $[T_{exp},0]$ to the interval $[-1,1]$.  
The coefficients and Chebyshev polynomials\cite{chebyshev} are presented in \cref{tab:coefficient}.
\begin{equation}
    \begin{split}
    Chebyshev(x) = &C_0T_0(x_t) + C_1T_1(x_t) + C_2T_2(x_t) + C_3T_3(x_t) + C_4T_4(x_t)\\ 
    &+ C_5T_5(x_t) + C_6T_6(x_t) + C_7T_7(x_t)
    \end{split}
    \label{eq:cheb}
\end{equation}

\begin{table}[t]
\setlength{\abovecaptionskip}{0cm}
  \caption{The coefficients and Chebyshev polynomials of exponential function.}
  \label{tab:coefficient}
  \centering
  \scalebox{0.8}{
  \begin{tabularx}{\textwidth}{l|X|X|X|X|X|X|X|X}
   \toprule   
    \midrule
    $i$ & 0 & 1 & 2 & 3 & 4 & 5 & 6 & 7 \\
    \hline
    $C_i$(0.) & 14021878 & 27541278 & 22122865 & 14934221 & 09077360 & 04369614 & 02087868 & 00996535 \\
    \hline
    $T_i(x)$ & 1 & $x$ & $2x^2-1$ & $4x^3-3x$ & $8x^4-8x^2+1$ & $16x^5-20x^3+5x$ & $32x^6-48x^4+18x^2-1$ & $64x^7-112x^5+56x^3-7x$\\
 \toprule
  \bottomrule
\end{tabularx}
}
\end{table}

 We also refer to Puma\cite{Puma} and replace the operation $\rm Div(\textbf{x}, Broadcast(\textit y))$ with $\rm \textbf{x}\cdot Broadcast( 1/\textit y )$ for further optimization, which ultimately saves computation and communication costs. The overall algorithm is demonstrated as \cref{alg:softmax}. First, multiple parties jointly compute $[\![\rm{\textbf{b}}]\!]^B$ and the maximum of $[\![\rm{\textbf{x}}]\!]$. Here $[\![\rm{\textbf{b}}]\!]^B = 1$ indicates that $[\![\rm{\textbf{x}}]\!]$ is greater than $ T_{exp}$, while $[\![\rm{\textbf{b}}]\!]^B = 0$ indicates the opposite. Next each party locally subtracts the maximum value from their respective $[\![\rm{\textbf{x}}]\!]$ and maps  resulting value to the interval $[-1,1]$.  This mapping transformation result is represented as $[\![\rm{\textbf{t}}]\!]$. Then they jointly compute the powers of $[\![\rm{\textbf{t}}]\!]$ and $[\![T_i(\rm{\textbf{t}})]\!]$ from 2 to 7, where $T_i(\textbf{t})$ represents the $i$-th order Chebyshev polynomial evaluated at $[\![\textbf{t}]\!]$. After that each party can compute the $Chebyshev([\![\rm{\textbf{x}}]\!])$, represented by  $[\![\rm{\textbf{z}}]\!]$, by summing the products of $C_j$ and $[\![T_j(\rm{\textbf{t}})]\!]$ from 0 to 7. The resulting $[\![\textbf{z}]\!]$ represents numerator of the SoftMax function. And $[\![z]\!]$, which is the sum of $[\![\textbf{z}]\!]$ across all parties, corresponds to denominator of the SoftMax calculation. Finally, parties compute and broadcast the reciprocal of $[\![\textbf{z}]\!]$, denoted as $[\![1/z]\!]$, and then compute $[\![\rm{\textbf{z}}/\textit z]\!]$. The SoftMax of $[\![\rm{\textbf{x}}]\!]$ can be obtained by multiplying $[\![\rm{\textbf{b}}]\!]^B$ and $[\![\rm{\textbf{z}}/\textit z]\!]$.

\begin{algorithm}[ht]
    \renewcommand{\algorithmicrequire}{\textbf{Input:}}
    \renewcommand{\algorithmicensure}{\textbf{Output:}}
    \caption{Secure SoftMax Protocol $\prod_{\rm{SoftMax}}$}
    \label{alg:softmax}
  \begin{algorithmic}[1]
  \REQUIRE{\ $P_i$ holds the 2-out-of-3 replicate secret share $[\![\rm{\textbf{x}}]\!]$ for $i \in \{0, 1, 2\}$}, and $\rm{\textbf{x}}$ is a vector of size $n$.
  \ENSURE{\ $P_i$ gets the 2-out-of-3 replicate secret share $[\![\rm{\textbf{y}}]\!]$ for $i \in \{0, 1, 2\}$, where $\rm{\textbf{y}} = \rm{SoftMax}(\rm{\textbf{x}})$.}
  
		\STATE $P_0, P_1, P_2$ jointly compute $[\![\rm{\textbf{b}}]\!]^B = \prod_{\rm{LT}}(T_{exp},[\![x]\!])$ and the maximum $[\![\Bar{x}]\!] = \prod_{\rm{Max}}([\![\rm{\textbf{x}}]\!])$.
        \STATE Locally compute  $[\![\Hat{\rm{\textbf{x}}}]\!] = [\![\rm{\textbf{x}}]\!] - [\![\Bar{x}]\!] - \epsilon$ and $[\![\rm{\textbf{t}}]\!] = -2*([\![\Hat{\rm{\textbf{x}}}]\!]-T_{exp})*T_{exp}^{-1}-1$.
        \FOR{i = 2, 3, ..., 7 }
            \STATE Jointly compute {$[\![\rm{\textbf{t}}^i]\!]=\prod_{\rm{Mul}}([\![\rm{\textbf{t}}^{i-1}]\!],[\![\rm{\textbf{t}}]\!])$} and $[\![T_i(\rm{\textbf{t}})]\!]$ based on $[\![\rm{\textbf{t}}^i]\!]$ as \cref{tab:coefficient}.
        \ENDFOR
        \STATE Locally compute $[\![\rm{\textbf{z}}]\!] = \sum_{j=0}^{7}C_j[\![T_j(\rm{\textbf{t}})]\!]$ and $[\![z]\!] = \sum_{k=1}^{n}[\![\rm{\textbf{z}}[\textit k]]\!]$.
        \STATE Jointly compute $[\![1/z]\!] = \prod_{\rm{Recip}}([\![z]\!])$ and $[\![\rm{\textbf{z}}/\textit z]\!]=\prod_{\rm{Mul}}([\![\rm{\textbf{z}}]\!], [\![1/\textit z]\!]) $.
        \STATE \textbf{return} $[\![\rm{\textbf{y}}]\!] = \prod_{\rm{Mul_{BA}}}([\![\rm{\textbf{b}}]\!]^B,[\![\rm{\textbf{z}}/\textit z]\!])$.
		
    \end{algorithmic}
  
\end{algorithm}

\subsection{Secure Activations}
\label{sec:activa}
ReLU, SiLU and Mish\cite{mish} are three main activation functions of U-Net\cite{unet} in DMs . ReLU, short for Rectified Linear Unit, is defined as $\rm{ReLU}(\textit x)=max(0,\textit x)$. It is a simple and computationally efficient activation function that involves only linear operations. In contrast, SiLU (Sigmoid-Weighted Linear Unit) and Mish\cite{mish} are activation functions that involve more complex non-linear operations. SiLU, also known as Swish, is defined as $\rm{SiLU}(\textit x) = \textit x\ast \rm{Sigmoid}(\textit x)$ where $\rm{Sigmoid}(\textit x)=\frac{1}{1+e^{-\textit x}}$. Mish, proposed by Diganta Misra\cite{mish}, is defined as $\rm{Mish}(\textit x)=\textit x\ast \rm{Tanh}(\rm{Softplus}(\textit x))$ where $\rm{Softplus}(\textit x)=ln(1+e^\textit x)$. To optimize the computational efficiency of exponential (exp) and hyperbolic tangent (tanh) functions, we replace them with linear piecewise fitting functions shown as \cref{eq:activation}. Considering both SiLU and Mish functions exhibit almost linear on the two sides (\ie $\rm{SiLU}/\rm{Mish}(\textit x)\approx 0$ for $x< -6$ and $\rm{SiLU}/\rm{Mish}(x)\approx x$ for $\textit x>6$), we perform the piecewise polynomial fitting at two intervals $[-6,-2],[-2,6]$ with maximum orders of 2 and 6 respectively. Besides, we only utilize even order terms for interval $[-2,6]$ to reduce the number of multiplications. Polynomials $F_0()$ and $F_1()$ are computed by library numpy.ployfit\footnote{https://numpy.org/doc/stable/reference/generated/numpy.polyfit.html} expressed as \cref{eq:silu} for SiLU and \cref{eq:mish} for Mish.

\begin{equation}
    \rm{Activation}(\textit x)=
    \begin{aligned}
    \begin{cases}
     0, &\ x<-6 \\
     F_0(x), &\  -6<=x<-2 \\
     F_1(x), &\  -2<=x<=6 \\
     x, &\  x>6
     \end{cases}
    \end{aligned}
    \label{eq:activation}
\end{equation}

\begin{equation}
    \left\{
    \begin{array}{ll}
    F_0(x)=&-0.01420163 x^2 -0.16910363 x -0.52212664\\
    F_1(x)=&0.00008032 x^6 -0.00602401 x^4 +0.19784596 x^2 
      + 0.49379432 x \\&+ 0.03453821\\
    \end{array}
    \right.
    \label{eq:silu}
\end{equation}

\begin{equation}
    \left\{
    \begin{array}{ll}
    F_0(x)=&-0.01572019 x^2 -0.18375535 x -0.55684445 \\
    F_1(x)=&0.00010786 x^6 -0.00735309 x^4  + 0.20152583 x^2 + 0.54902050 x \\
    &+ 0.07559242\\
    \end{array}
    \right.
    \label{eq:mish}
\end{equation}

Based on above fitting piecewise polynomials, we design secure Activation Protocol as  \cref{alg:seact}. First, parties jointly computes $[\![z_0]\!]^B, [\![z_1]\!]^B, [\![z_2]\!]^B$ which represent the location interval of $[\![x]\!]$. Then they compute $[\![x^2]\!], [\![x^4]\!], [\![x^6]\!]$ by $\prod_{\rm{Square}}$ and $\prod_{\rm{Mul}}$ protocol. $[\![F_0(x)]\!]$ and $[\![F_1(x)]\!]$ can be easily computed in local. Finally they obtain the SiLU or Mish function by $\prod_{\rm{Mul_{BA}}}$ protocol and add each part.

\begin{algorithm}[ht]
    \caption{Secure SiLU and Mish Protocol $\prod_{\rm{SiLU}}$/$\prod_{\rm{Mish}}$}
    \label{alg:seact}
    \renewcommand{\algorithmicrequire}{\textbf{Input:}}
    \renewcommand{\algorithmicensure}{\textbf{Output:}}
  \begin{algorithmic}[1]
  \REQUIRE{\ $P_i$ holds the 2-out-of-3 replicate secret share $[\![x]\!]$ for $i \in \{0, 1, 2\}$}.
  \ENSURE{\ $P_i$ gets the 2-out-of-3 replicate secret share $[\![y]\!]$ for $i \in \{0, 1, 2\}$, where $y = \rm{SiLU}(x)/\rm{Mish}(x)$.}

		\STATE $P_0, P_1, P_2$ jointly compute $[\![b_0]\!]^B, [\![b_1]\!]^B, [\![b_2]\!]^B$
        and $[\![z_0]\!]^B, [\![z_1]\!]^B, [\![z_2]\!]^B$ where 
            $$
            \begin{array}{lll}
            [\![b_0]\!]^B=\prod_{\rm{LT}}([\![x]\!],-6), &\quad [\![b_1]\!]^B=\prod_{\rm{LT}}([\![x]\!],-2), &\quad [\![b_2]\!]^B=\prod_{\rm{LT}}(6, [\![x]\!]), \\

            [\![z_0]\!]^B = [\![b_0]\!]^B \oplus [\![b_1]\!]^B, & \quad [\![z_1]\!]^B = [\![b_1]\!]^B \oplus [\![b_2]\!]^B \oplus 1, & \quad [\![z_2]\!]^B = [\![b_2]\!]^B.
            \end{array}
            $$
        Note that $z_0=1\{-6\leq x<-2\}, z_1 = 1\{-2\leq x\leq 6\}$ and $z_2 = 1\{x>6\}$.
            
        \STATE Jointly compute $[\![x^2]\!]=\prod_{\rm{Square}}([\![x]\!])$, $[\![x^4]\!]=\prod_{\rm{Square}}([\![x^2]\!])$, and $[\![x^6]\!]=\prod_{\rm{Mul}}([\![x^2]\!], [\![x^4]\!])$.
        
        \STATE Locally compute polynomials $[\![F_0(x)]\!]$ and $[\![F_1(x)]\!]$ based on $[\![x^i]\!]$ as \cref{eq:silu} / \cref{eq:mish}.
        \STATE \textbf{return} \\$[\![y]\!] = \prod_{\rm{Mul_{BA}}}([\![z_0]\!]^B, [\![F_0(x)]\!]) + \prod_{\rm{Mul_{BA}}}([\![z_1]\!]^B, [\![F_1(x)]\!])$$ + \prod_{\rm{Mul_{BA}}}([\![z_2]\!]^B, [\![x]\!])$.
    \end{algorithmic}
\end{algorithm}

\subsection{Secure Time Embedding}
\label{sec:temb}
The time embedding module (tMLP) comprises one Positional Embedding layer, two Linear layers and one SiLU layer. However, the exponentiation operation in both Positional Embedding layer and  SiLU layer contributes to most latency. To reduce running time we replace the exponentiation in Positional Embedding layer with Chebyshev fitting as \cref{eq:cheb}, and compute SiLU layer by $\prod_{\rm{SiLU}}$.

\section{Experiments}
\textbf{Implementation.} We implement CiperDM on top of SPU\cite{spu} in C++ and Python. SPU compiles a high-level Flax code to secure computation protocols, which are then executed by designed cryptographic backends. We run our experiments on Ubuntu 20.04.1 LTS with Linux kernel 5.4.0-146-generic. The CPU mode is Intel(R) Xeon(R) Silver 4314 CPU @2.40GHz with 500GB RAM and a single thread. We use Linux tc tool by Cheetah\cite{Cheetah} to simulate local-area network (LAN, RTT: 0.1 ms, 1 Gbps). All frameworks are measured in both local and LAN scenes.\\
\textbf{Models \& Datasets.} We evaluate CipherDM on diffusion model architectures: DDPM, DDIM and SD. We measure the sampling performance for DDPM and DDIM over MNIST dataset. As models of SD pipeline are too large to be deployed under SPU, we only utilize the U-Net\cite{unet} module for assessment.\\
\textbf{Baseline.} We compare CipherDM with the direct implementation on SPU.

\subsection{Inference Costs}
In this subsection, we conduct experiments to obtain inference costs on CPU, SPU, and our CipherDM framework. Sampling costs of CipherDM  are compared with direct deployment on SPU. We perform  evaluation on MNIST dataset using  DDPM and DDIM methods. Specifically, we measure the costs associated with sampling one single image using ReLU, SiLU, and Mish activations  individually. The image size is (28,28) , the batchsize is 1, and the data type is float32. We set 1000 sampling steps for DDPM, whereas DDIM just needs 50 steps to simulate this. Time is in seconds and communication (Comm. for short) is in GB.

\begin{table}[t]
\setlength{\tabcolsep}{1mm}
  \caption{\small Total time costs of sampling one single image. }
  \label{tab:time}
  \centering
  \scalebox{0.8}{
  \begin{tabular}{c c|c c|c  c|c  c}
   \toprule   
    \midrule 
    \multicolumn{2}{c|}{\multirow{2}{*}{Framework}}&  \multicolumn{2}{c|}{ReLU} & \multicolumn{2}{c|}{SiLU} & \multicolumn{2}{c}{Mish} \\
   
    & & Local & LAN &Local & LAN & Local & LAN\\
    \hline
    \multirow{4}{*}{DDPM}&CPU&373 & 377   & 370 &369  &372 & 376\\
    
     &SPU &  9879 & 17265 & 13054 & 24042& 17698 & 31002\\
    % \cline{2-11}
    &CipherDM& 9473& 16669& 9688& 17915 & 9520&16686\\
    
    &\cellcolor{black!10} Improv. &\cellcolor{black!10} 1.043$\times$ &\cellcolor{black!10} 1.037$\times$ &\cellcolor{black!10} 1.347$\times$ &\cellcolor{black!10} 1.342$\times$ &\cellcolor{black!10} 1.859$\times$ &\cellcolor{black!10} 1.858$\times$\\
    \hline
    \multirow{4}{*}{DDIM}&CPU& 28& 27   & 28 & 27 & 28& 28\\
    
    &SPU &  587   &949 & 808 & 1250 & 1047&  1781\\
    
    &CipherDM&534  & 910& 648 & 1088 & 576 &979 \\
    
    &\cellcolor{black!10} Improv.  &\cellcolor{black!10}1.099$\times$  &\cellcolor{black!10} 1.043$\times$  &\cellcolor{black!10} 1.247$\times$  &\cellcolor{black!10} 1.241$\times$ &\cellcolor{black!10} 1.818$\times$  &\cellcolor{black!10} 1.819$\times$\\
 \toprule
  \bottomrule
\end{tabular}
 }
 
\end{table}
\cref{tab:time} reveals that
CipherDM outperforms SPU in terms of speed for the entire pipeline. Specifically, for DDPM, CipherDM is approximately $1.037\times \sim 1.859 \times$ faster than SPU. Similarly, for DDIM, CipherDM expresses a speed advantage ranging from $1.043\times$ to $1.819\times$ faster than SPU. 

\begin{table}[ht]
\setlength{\tabcolsep}{1mm}
  \caption{\small MPC time and communication costs of  sampling one single image. }
  \label{tab:cost}
  \centering
  \tabcolsep=0.1cm
  \scalebox{0.8}{
  \begin{tabular}{c c|c c|c c|c c}
   \toprule   
    \midrule 
\multicolumn{2}{c|}{\multirow{2}{*}{Framework}}
      & \multicolumn{2}{c|}{ReLU} & \multicolumn{2}{c|}{SiLU} & \multicolumn{2}{c}{Mish} \\

    & & Time & Comm. & Time & Comm. & Time & Comm. \\
    \hline
    \multirow{3}{*}{\shortstack{Local\\(DDPM)}} & SPU & 7076& 186.18 & 9794  & 268.04&14295& 391.11\\
    
    & CipherDM & 6069& 198.61 & 7278  &198.72 &6405 & 230.33\\
    
    & \cellcolor{black!10}Improv.&\cellcolor{black!10} 1.166$\times$ &\cellcolor{black!10} 0.937$\times$ &\cellcolor{black!10} 1.346$\times$  &\cellcolor{black!10}1.349$\times$ &\cellcolor{black!10}2.232$\times$ &\cellcolor{black!10} 1.698$\times$ \\
    \hline
    \multirow{3}{*}{\shortstack{LAN\\(DDPM)}} & SPU & 14816& 186.10 & 20679 &268.04 &27586 & 391.07\\
    
    & CipherDM & 12174& 198.40 & 14583  & 198.84&12660 & 230.31\\
    
    & \cellcolor{black!10}Improv.&\cellcolor{black!10}1.217$\times$ & \cellcolor{black!10}0.938$\times$ &\cellcolor{black!10}  1.418$\times$ &\cellcolor{black!10} 1.348$\times$&\cellcolor{black!10} 2.179$\times$&\cellcolor{black!10} 1.698$\times$ \\
    \hline
    \multirow{3}{*}{\shortstack{Local\\(DDIM)}} & SPU & 492& 9.13 & 712  &13.42 &957 &19.84 \\
    
     & CipherDM & 413&10.00  & 474  &11.07 &411 & 11.07\\
    
    & \cellcolor{black!10}Improv.&\cellcolor{black!10}1.191$\times$ &\cellcolor{black!10} 0.915$\times$ &\cellcolor{black!10} 1.502$\times$  &\cellcolor{black!10}1.212$\times$ &\cellcolor{black!10}2.328$\times$ &\cellcolor{black!10} 1.791$\times$ \\
    \hline
    \multirow{3}{*}{\shortstack{LAN\\(DDIM)}} & SPU & 855& 9.13 & 1260  & 13.44&1678 & 19.85\\
    
    & CipherDM & 789& 9.56 & 907  & 11.08&795 &11.09 \\
    
    & \cellcolor{black!10}Improv.&\cellcolor{black!10}1.084$\times$ &\cellcolor{black!10}0.917$\times$  &\cellcolor{black!10} 1.389$\times$  &\cellcolor{black!10} 1.213$\times$&\cellcolor{black!10}2.111$\times$ &\cellcolor{black!10} 1.790$\times$ \\

 \toprule
  \bottomrule
\end{tabular}
}
 
\end{table}

As depicted in \cref{tab:cost}, 
we can see that when considering the execution of MPC alone, CipherDM demonstrates faster computation and improved communication efficiency compared to SPU. For DDPM, CipherDM is approximately $1.166\times \sim 2.232 \times$ faster in terms of computation speed, and it is $1.348\times \sim 1.698 \times$ more communication-efficient compared to SPU. Similarly, for DDIM, CipherDM achieves a speed advantage of approximately $1.084\times \sim 2.328 \times$ and is $1.212\times \sim 1.791 \times$ more communication-efficient than SPU.

It's worth noting that in CipherDM, the communication cost for ReLU is comparatively higher than other activations. This results from the secure SoftMax block, which reduces running time but increases communication requirements. Overall, CipherDM offers faster computation and improved communication efficiency for both DDPM and DDIM, making it a favorable choice in practical scenarios.

\begin{table}[t]
    \centering
    \begin{minipage}{0.45\textwidth}
        \centering
        \caption{ FID of 10k images generated  by CPU and CipherDM in plaintext.}
        \label{tab:fid}
      \tabcolsep=0.2cm
      \scalebox{0.8}{
      \begin{tabular}{c|c|c|c }
       \toprule   
        \midrule 
         FID & ReLU & SiLU & Mish   \\
        \hline
        CPU & 110.13 & 79.46 & 86.24  \\
        % \hline
        CipherDM&272.26 & 252.12 & 202.64   \\
     \toprule
      \bottomrule
    \end{tabular}
    }
    \end{minipage}
    \hspace{0.02\textwidth}
    \begin{minipage}{0.5\textwidth}
        \centering
        \caption{ Time costs of U-Net in Flax Stable Diffusion from Diffusers\cite{diffusers}.}
        \label{tab:sdunet}
      \tabcolsep=0.2cm
      \scalebox{0.8}{
      \begin{tabular}{c|c|c|c|c }
       \toprule   
        \midrule 
         Numsteps & CPU & SPU & CipherDM & Improv.  \\
        \hline
        1& 44 & 8332& 7711 & 1.081$\times$  \\
        5& 81& 8856 & 8208 & 1.079$\times$  \\
     \toprule
      \bottomrule
    \end{tabular}
    }
    \end{minipage}  
\end{table}

\subsection{Accuracy \&  Image Quality}
\begin{figure*}[t]
    \centering
    \begin{minipage}[t][3.1cm]{0.45\textwidth}
       \centering
        \includegraphics[width=\textwidth]{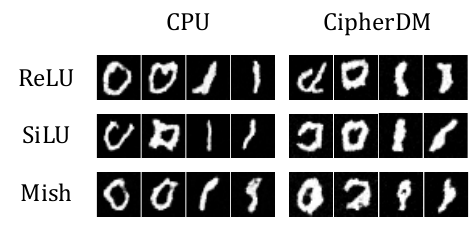} 
        \caption{Images generated  by CPU and CipherDM.}
        \label{fig:01}
    \end{minipage}
    \hfill
    \begin{minipage}[t][3.1cm]{0.45\textwidth}
        \centering
        \includegraphics[width=\textwidth]{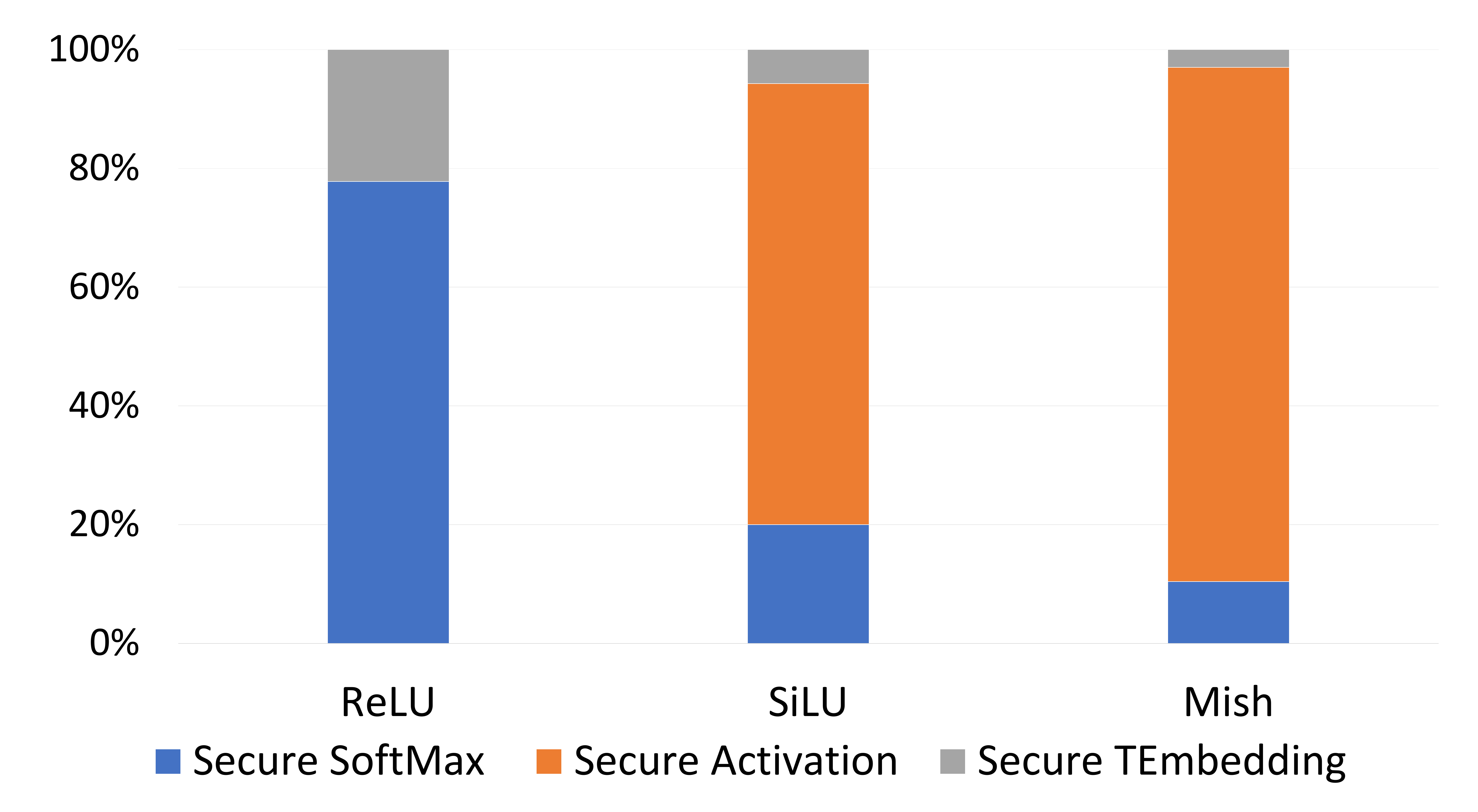}
  \caption{ The impact of each module on total time improvement.
  }
  \label{fig:percentage}
    \end{minipage}
\end{figure*}

 Given the slow generation in secret mode, we 
 evaluate the accuracy and image quality by comparing the images generated by CPU and CipherDM in plaintext. The MSE of our approximation in single activation falls within the range of $10^{-6}\sim 10^{-3}$, which is negligible. Besides, the generated images presented in  \cref{fig:01} show no obvious quality decay. \cref{tab:fid} displays the Fréchet Inception Distance (FID) of 10k generated images in plaintext. A lower FID score signifies better alignment between the generated samples and real images. We observe that some generated images contain substantial noise, while others appear normal, which may explain the FID decline.

\subsection{Evaluation on Stable Diffusion}
Initially, our intention was to evaluate CipherDM on the complete Stable Diffusion pipeline. However, due to the large size of many models involved, including  text encoder, VAE, and U-Net, executing the pipeline under SPU became challenging due to the limitation of its underlying mechanism. Therefore, we focused our measurements solely on the U-Net component, which serves as the central core of Stable Diffusion.

The results presented in \cref{tab:sdunet} demonstrate that CipherDM can significantly enhance running time, with improvement of up to approximately $1.081\times$ compared to the baseline execution without CipherDM. Although we were unable to evaluate  complete Stable Diffusion pipeline, these findings highlight the potential of CipherDM in accelerating the U-Net component, which plays a crucial role in overall Stable Diffusion process.
\subsection{Effect of Each Protocol}
 During the denoising process of DMs, noise prediction is performed using a network, typically U-Net. This procedure is replicated multiple times. We measure and analyze the impact of each protocol during a single execution of U-Net. 
 
 The results presented in \cref{fig:percentage} eval the contribution of different protocols to  the overall improvement.  When using ReLU activation, the Secure SoftMax protocol accounts for approximately 77.8\% of total improvement, while the Secure Time Embedding (TEmbedding) protocol contributes about 22.2\%. However, when SiLU or Mish activations are employed, the Secure Activation protocol becomes more significant, contributing 74.3\% and 86.6\% respectively. 
 
Additionally, it is important to note that the time embedding module only requires one execution for entire sampling process, while the SoftMax and activation modules are executed proportionally to the number of sampling steps. Therefore, as the number of sampling steps increases, the impact of the Secure SoftMax and Secure Activation protocols becomes more pronounced, while the influence of the Secure Time Embedding protocol diminishes.

\section{Conclusion}
In this paper, we introduce CipherDM, a novel MPC framework designed to address privacy concerns in secure sampling on Diffusion Models. Our framework aims to guarantee secure sampling with MPC technology and improve the efficiency by approximating computationally expensive activation functions with accurate polynomials. Additionally, we propose secure protocols for SoftMax and SiLU/Mish activations. While the current sampling efficiency may not be practical, this work represents an important initial step towards addressing privacy issues associated with Diffusion Models. Further work will focus on enhancing computational efficiency of secure sampling by modifying the model architecture and designing more efficient underlying protocols. By combining CipherDM with other optimization methods and leveraging hardware acceleration, we envision that secure Diffusion Model sampling will become applicable in practical scenarios in the future.

% \section*{Acknowledgement}
% This work was supported in part by the Beijing Municipal Science Technology Commission New generation of information and communication technology innovation Research and demonstration application of key technologies for privacy protection of massive data for large model training and application (Z231100005923047).

% \clearpage  % TODO REVIEW/FINAL: This \clearpage needs to be removed from both review and camera-ready versions.
% \bibliography{egbib}

\end{document}